\documentclass[aps,pra,twocolumn,superscriptaddress]{revtex4-1}
\usepackage{graphicx}
\usepackage{amsmath,amssymb,amsfonts}
\usepackage{color}

\begin{document}

%\setpagewiselinenumbers
%%\modulolinenumbers[1]
%\linenumbers

%\title{Electron microscopy diffraction modeling using quantum trajectories}
\title{A novel quantum dynamical approach in electron microscopy\\
combining wave-packet propagation with Bohmian trajectories}

\author{S. Rudinsky}
\affiliation{Department of Mining and Materials Engineering, McGill
University, 3610 University Street, Montreal, Qc, Canada, H3A 0C5}

\author{A. S. Sanz}
\affiliation{Department of Optics, Faculty of Physical Science,
Universidad Complutense de Madrid,\\
Pza.\ Ciencias 1, 28040 Madrid, Spain}

\author{R. Gauvin}
\affiliation{Department of Mining and Materials Engineering, McGill
University, 3610 University Street, Montreal, Qc, Canada, H3A 0C5}

\begin{abstract}
The numerical analysis of the diffraction features rendered by
transmission electron microscopy (TEM) typically relies either on
classical approximations (Monte Carlo simulations) or quantum paraxial
tomography (the multislice method and any of its variants).
Although numerically advantageous (relatively simple implementations
and low computational costs), they involve important approximations and
thus their range of applicability is limited.
To overcome such limitations, an alternative, more general
approach is proposed, based on an optimal combination of wave-packet
propagation with the on-the-fly computation of associated Bohmian
trajectories.
For the sake of clarity, but without loss of generality, the approach
is used to analyze the diffraction of an electron beam by a thin
aluminum slab as a function of three different incidence (work)
conditions which are of interest in electron microscopy: the probe width, the
tilting angle, and the beam energy.
Specifically, it is shown that, because there is a dependence on
particular thresholds of the beam energy, this approach provides a
clear description of the diffraction process at any energy, revealing
at the same time any diversion of the beam inside the material towards
directions that cannot be accounted for by other conventional methods,
which is of much interest when dealing with relatively low energies
and/or relatively large tilting angles.
\end{abstract}

\date{\today}

%\pacs{}

\maketitle

%%%%%%%%%%%%%%%%%%%%%%%%%%%%%%%%%%%%%%%%%%%%%%%%%%%%%%%%%%%%%%%%%%%%%%%

\section{\label{sec1} Introduction}
Characterization through electron microscopy (EM) is an important
aspect in materials development. Microscopic features and
crystallographic information can be elucidated from high
resolution and electron diffraction imaging
\cite{williams2009transmission}. While much information can be
obtained from diffraction imaging, data analysis is complex and must
be coupled with simulations in order to be properly utilized.
Currently, two algorithms are used to simulate selected-area
diffraction patterns in conventional TEM (CTEM): the Bloch wave
method and the multislice method, both time-independent approaches
to image simulations \cite{kirkland2010advanced}. The Bloch wave
method consists in solving for the coefficients of each Bloch wave
given the specific periodicity of the crystal. Recently, Mitsuichi
{\it et al.}\ \cite{Mitsuishi2008981} have developed a routine
implementing this computational method for scanning confocal EM.
Routines using Bloch wave analysis have also been developed
\cite{Pennycook1990PhysRevLett,Watanabe2001PhysRevB} for angular
dark-field scanning TEM (STEM) imaging.
Because computational times increase exponentially when more than
two beams are used to calculate the Bloch wave expression, the
multislice method has become the technique more commonly used for
diffraction simulations \cite{kirkland2010advanced}. The material is
separated into slices perpendicular to the incident beam and the
wave function is calculated iteratively over each slice.
This procedure takes advantage of the paraxiality arising from the
relatively large energies typically involved in TEM.
Consequently, a lot of codes implementing the multislice approach have
been developed and are available in the literature.
Notably this has been done by Kirkland for CTEM and STEM at
accelerating voltages greater or equal to 100 keV
\cite{kirkland2010advanced,KIRKLAND198777}.
Recently, G\'omez-Rodr{\'\i}guez {\it et al.}\
\cite{GomezRodriguez201095} have also developed a routine based on the
multislice approach to tackle amorphous materials.

As mentioned above, the multislice algorithm is based on the paraxial
approximation, which relies on the assumption of fast (highly energetic)
electrons and small (negligible) scattering angles
\cite{kirkland2010advanced}, thus facilitating computations and
decreasing computing times.
This approximation is acceptable in TEM, where accelerating voltages
are typically within the range of 100-200~keV.
However, it is no longer applicable at lower
voltages, such as those used in a scanning electron microscope (SEM),
where backscattering and high scattering angles become significant
\cite{goldstein2003scanning}.
Furthermore, another drawback of diffraction
simulation methods thus far is the inability to reproduce effects
that result from the particle characteristics of electrons. These
may include secondary and backscattering coefficients, electron
transport and inelastic scattering \cite{goldstein2003scanning}.
Monte Carlo simulation methods are currently used in that effect.
Work has been performed by Gauvin {\it et al.}\
\cite{Gauvin1995Scanning,gauvin2006win} in characterizing
electron diffusion processes using probabilistic methods based on
random number generators which provide quantitative information
about particle collision processes.
Lloevet {\it et al.}\ \cite{Llovet2004} have also employed these
methods for electron microanalysis of bulk specimens.
Overall, electron transport is primarily simulated with a multitude of Monte
Carlo algorithms which compute a range of parameters and
coefficients that describe the EM
system \cite{joy1995monte,echlin2013advanced}. However, these
calculations only take into account the particle-like nature of
electrons and use general probability models associated with
classical trajectories.

There is therefore a crucial need in
the field of EM to describe, explain, understand,
and interpret observations made at probing conditions outside of
those used in conventional TEM.
This has generated much interest in the implementation and
development of new methods of image simulation that fulfill such
a need.
The goal of the present work goes in this direction.
Here an alternative, more general approach is introduced, which
optimally combines standard wave-packet propagation with the on-the-fly
computation of associated Bohmian trajectories\cite{sanz2012trajectory}
to produce EM simulations that circumvent the drawbacks of both the
multislice and Bloch wave algorithms, on the one hand, and classical
Monte Carlo simulations, on the other hand.
As a result, this approach is able to provide time-dependent information about
the system that can be used to incorporate particle aspects into EM,
unifying both the wave and particle-like natures of electrons within
a single working framework.
This is possible, because the Bohmian formulation of quantum
mechanics has the advantage of describing quantum phenomena by means
of a hydrodynamic language, where the evolution of the quantum system
is monitored by means of individual paths or trajectories that do not
contravene any of the fundamental principles of the quantum theory.
Any outcome (quantum observable) is then reproduced by statistically
analyzing the behavior of swarms of such trajectories, which evolve
under the guidance of the wave function.

A full characterization of the different dynamical regimes that can be
observed in diffraction of rare gases by metal surfaces at low incident
energies, namely the gradual transition from the Fresnel to the
Fraunhofer diffraction, has already been reported by means of this
technique \cite{sanz:prb:2000,Sanz2002Diff}, as well as the effects of
increasingly more massive probes \cite{sanz:EPL:2001} or turbulence
(vortical dynamics) arising under presence of impurities on the surface
\cite{sanz:jcp:2004,sanz:prb:2004}.
Similar quantitative analyses of grating diffraction of low energy
electrons, neutrons, or fullerenes have also been reported in the
literature \cite{Sanz2002Diff,sanz:SSR:2004}, finding interesting
analogies between the behavior displayed by the diffracted system and
manifestations found in other different physical contexts, such as the
relationship between the formation of Talbot carpets and wave-guiding
or Bloch-like periodic invariance \cite{Sanz2007}.
More recently, these kind of applications have also been extended to the
field of EM by Zhang {\it et al.} \cite{Zhang2015JMicro},
which typically involves much higher energies.
In particular, these authors analyzed STEM with trajectories obtained
from the multislice method, which in virtue of the paraxial approximation
enabled by high energies allows to reparameterize the forward coordinate
in terms of time, thus producing an effective reduction of the problem
dimensionality, from three to two dimensions (namely the transversal
ones).
As mentioned above, this means that the method gains some efficiency
regarding computational cost, but undergoes the same shortcomings of
the multislice method, thus not providing us with any additional
insight into electron-material interactions, such as backscattering
and high-angle collisions.
To overcome this inconvenience, the strategy followed in this work
consists in computing the trajectories on-the-fly, as the wave function
evolves, but without reducing the dimensionality of the problem, just
following a standard wave-packet propagation scheme.
In order to gain some efficiency and ensure numerical stability,
the guidance equation takes advantage of the spectral decomposition
employed to propagate the wave function \cite{sanz2014trajectory}.
To show the feasibility of the proposed methodology, we have analyzed
the problem of electron beam diffraction by a thin Al[100] slab as a
function of a series of incident probing conditions of experimental
interest, such as the probe width ($\sigma_0$), the tilt angle
($\theta_i$), and the incidence energy ($E_i$).

The remainder of this work has been organized as follows. In the next
section, we describe the modeling of the problem, including the
algorithm used to obtain quantum trajectories from the wave function
as well as the scattering potential model used to reproduce the target,
namely a thin Al film formed by several atom layers.
For simplicity, but without loss of generality, we have
decided to consider as a working model a dimensionally reduced crystal.
This allows us to focus on and describe more clearly the electron beam
dynamics along the perpendicular and parallel directions with respect
to the surface of the crystal slab.
Specifically, this reduced model represents the projection of a
face-centered cubic (FCC) lattice with a thickness of about 80~\AA\
(around 20 unit cell layers).
Regarding the wave function simulation, an incident probe beam represented by a Gaussian wave packet has been considered, with the associated Bohmian
trajectories launched from a series of initial positions distributed in
a way that they are able to map the propagation of each portion
of the wave function. This provides specific information about
how the beam spreads throughout the material and afterwards in order to understand and explain the role played by the incidence conditions
in the diffraction process.
In Sec.~\ref{sec3}, we present and discuss the main results obtained
from our simulations with the different probing conditions
specified above to investigate the response of the diffracted system
inside and outside the material under each parameter.
Finally, the main conclusions from this work are summarized in
Sec.~\ref{sec4}.

%%%%%%%%%%%%%%%%%%%%%%%%%%%%%%%%%%%%%%%%%%%%%%%%%%%%%%%%%%%%%%%%%%%%%%%

\section{\label{sec2} The Model}

Typically the fundamental equations involved in Bohmian mechanics
are obtained by substituting into the non-relativistic
Schr\"{o}dinger equation for a particle of mass $m$,
\begin{equation}
 \label{eq:schrodinger}
 i\hbar\frac{\partial \Psi}{\partial t} =
 -\frac{\hbar^2}{2m}\ \! \nabla^2\Psi + V\Psi ,
\end{equation}
the wave function in polar form, $\Psi=\sqrt{\rho}e^{iS/\hbar}$,
and then splitting the real and imaginary parts of the resulting
equation.
The two coupled differential equations that follow from this nonlinear
transformation (from a complex field variable to two real field
variables) correspond to the usual continuity equation, for $\rho$, and
a Hamilton-Jacobi one, for $S$, respectively.
It is through the latter equation that Bohm postulated
\cite{bohm:PR:1952-1} the possibility to map the evolution of the
quantum system in terms of trajectories developing in the corresponding
configuration space, which would follow a guidance equation
analogous to the classical Jacobi law.
Nonetheless, there is no need to postulate such an equation of motion,
since it also arises from a hydrodynamic approach to quantum mechanics.
Specifically, the quantum continuity equation reads as
\begin{equation}
 \frac{\partial \rho}{\partial t} + \nabla \cdot {\bf J} = 0 ,
\end{equation}
where
\begin{equation}
\label{eq:flux}
 J = \frac{\hbar}{2mi}
  \left(\Psi^*\nabla\Psi-\Psi\nabla\Psi^*\right)
\end{equation}
is the usual quantum current density or quantum flux \cite{schiff-bk}.
Taking into account the fact that in fluid dynamics the second term
can be recast as the product of the probability density $\rho$ and a
certain flow velocity vector field ${\bf v}$, i.e.,
${\bf J} = \rho {\bf v}$, we get
\begin{equation}
 \label{eq:bmmotion}
 {\bf v} = \dot{\bf r} = \frac{\bf J}{\rho} = \frac{\nabla S}{m} .
\end{equation}
This equation of motion describes the transport of the probability
density throughout configuration space in the form of quantum
streamlines or trajectories, which are generally known as Bohmian
trajectories.

It is worth noting that the paraxial approximation used by the
multislice algorithm follows from Eq.~(\ref{eq:schrodinger}) under the
condition
\begin{equation}
 \label{eq:msapprox}
 \left|\frac{\partial^2\Psi}{\partial z^2}\right|
  \ll \left|\frac{1}{\lambda}\frac{\partial \Psi}{\partial z}\right| ,
\end{equation}
where $\lambda$ is the electron wavelength and $z$ is the perpendicular
direction with respect to the surface of the crystal slab.
This approximation allows the recasting of the time-dependent Schr\"odinger
equation in terms of a dimensionally reduced $z$-dependent
Schr\"{o}dinger wave function that evolves parametrically along the
parallel direction as a function of the $z$ coordinate (keeping a
linear relationship with the propagation time).
At 100~keV, this approximation has been shown to yield accurate
results for image simulations at the Bragg angle, which is
considerably small for such a high accelerating voltage
\cite{HowiePhylMag1968}. If the Bragg angle is small, the tilt can
be represented as a small momentum contribution perpendicular to the
incident direction without including any rotation
\cite{kirkland2010advanced}. However, at low accelerating voltages,
such approximations cannot be made, hence the need for an alternative
simulation method.

Equation~(\ref{eq:bmmotion}) has been a source for a series of
quantum methodologies using the Bohmian trajectory as the fundamental
element \cite{trahan2006quantum}.
However, it can also be used together with the wave function to
generate other alternative schemes, where the trajectory does not
become the numerical solver, but an additional variable to investigate
the evolution of the quantum system in a statistical fashion.
In this latter case, a two-step process is typically followed to
obtain the numerical solution of Eq.~(\ref{eq:bmmotion}) at each
time step, generating in a recursive fashion the corresponding Bohmian
trajectories.
The approach here follows this scheme.
In particular, first the time-dependent Schr\"{o}dinger equation
is numerically solved by means of the split-operator method
\cite{FEIT1982412,feit-fleck:JCP:1983,kosloff:JCP:1983,kosloff:JCompPhys:1983}.
This equation can be solved by a variety of numerical methods available
in the literature (see, for instance,
Refs.~\onlinecite{leforestier:jcompphys:1991,zhang-bk,tannor2007introduction},).
The split-operator method has been chosen here, because it properly
matches the on-the-fly computation of the trajectories according to the
propagation scheme proposed in \cite{sanz2014trajectory} (see Appendix~A.3).
This takes advantage of the plane-wave Fourier decomposition of the wave function
at each time to recast Eq.~(\ref{eq:bmmotion}) in terms of an analytical
function, thus reducing natural propagation errors that arise when
the latter equation is fully numerically solved.
Furthermore, the integration of the Bohmian trajectories requires a
relatively small time-step (smaller, in general, than the steps that
can be considered for only the propagation of the wave function), and
therefore does not benefit from other large-step methods (e.g., the
Chebyshev or the Lanczos propagation schemes
\cite{leforestier:jcompphys:1991}).

The split-operator algorithm requires at least four
Fourier transforms per time step (forward and backward, twice), so to
enhance the efficiency of the process the fast Fourier transform (FFT)
algorithm \cite{press-bk-2} has been implemented.
The propagation of the wave function has been carried out on a
1024$\times$1024 grid, which in spite of its large size is appropriate
for the purposes of this work.
For more realistic three-dimensional simulations, however, the grid
size has to be remarkably reduced in order to gain efficiency and save
computation time (some tests are currently on course in this regard).
A large size for the grid has been considered here in order to better
understand the electron dynamics without the presence of absorbing
boundaries. To avoid the possibility that reflection ripples from the boundaries could
affect the whole of the transmitted wave function, a careful control
over the propagation time was taken by analyzing the behavior of a
series of restricted probabilities \cite{sanz:JPA:2011},
\begin{equation}
 \mathcal{P}_\Omega (t) = \int_\Omega |\Psi({\bf r},t)|^2 d{\bf r} ,
 \label{rest}
\end{equation}
where $\Omega$ refers to the corresponding region.
In particular, here we have established three regions limited by the
position of the Al slab: the incidence region (I), before the slab;
the internal region (II), inside the slab; and the transmission
region (III), outside the slab.
Thus, for each set of incidence parameters (particularly the incidence
energy), the simulations were always halted once the probability inside
the material was negligible and the whole of the wave function was
transmitted, but without displaying any significant reflection at the
boundaries.
Of course, for much lower energies than those explored here, absorbing
boundaries would have been used, since test calculations show how the
wave function flows towards the boundaries even inside the material
(due to internal diffraction parallel to the surface).

Then, with the plane-wave basis set provided by the FFT algorithm and
the corresponding Fourier coefficients, not only the local values of the
wave function can be determined everywhere throughout the numerical
grid (even outside grid points), but also, as mentioned above,
Eq.~(\ref{eq:bmmotion}) becomes an analytical function of the local
position and time, which allows us, in a second step, to use relatively
simple ordinary differential equation solvers, such as the Runge-Kutta
algorithm.
In our case, we have observed that a simple second-order Runge-Kutta
algorithm warrants numerical stability and accuracy comparing with the
results rendered by higher-order Runge-Kutta algorithms, which reduces
the computation time and enhances the efficiency of the method.
This on-the-fly approach remains still valid if instead of FFT
components one considers any general basis set from the momentum space
\cite{Sanz2007}.

In regards to the potential model describing the interaction between the
incoming electrons and the material, i.e., the potential $V$ in
Eq.~\ref{eq:schrodinger}, we have considered the functional form
proposed by Peng \cite{peng1999electron} for Al[100].
This model is a pairwise potential involving the distance between the
electron position ($r$) and the position of each lattice atom, $r_i$,
\begin{equation}
 \label{eq:sumpot}
 V(r)=\sum_i\phi_i(r-r_i) .
\end{equation}
Here $\phi_i$ denote the scattering potentials obtained by inverse
Fourier transforms of the elemental scattering factors,
\begin{eqnarray}
 \phi_i(r) & = & \frac{4\pi}{a^2}\left(\frac{\hbar^2}{m_0}\right)
 \sum_k a_k\ \! \sqrt{\frac{\pi}{b_k + B}} \nonumber \\
  & & \times \exp \left[-\frac{(b_k+B)G^2}{(4\pi)^2}
  -\frac{4\pi^2}{b_k+B}(r-r_i)^2 \right] ,
 \label{eq:scatpot}
\end{eqnarray}
where the constants $a_k$ and $b_k$ were obtained by Peng from a
fitting to the experimental data (see Table~\ref{tab:pengvalues}),
$a$ is the lattice parameter, $G$ is the modulus of the reciprocal
lattice vector, and $B$ is the elemental Debye-Waller factor.
The unit cell configuration of the material emulates the projection of
a face-centered cubic structure with 20 atomic layers, that is, each
atomic layer is shifted horizontally by half the lattice parameter
compared to the previous layer.
The Debye-Waller factor is introduced to phenomenologically include
the effects of thermal diffuse scattering, which for Al at 293~K is
0.7806 \cite{Peng:zh0008}.
As mentioned and explained above, we have reduced the dimensionality of
the problem, which means that we have considered a two-dimensional
projection of the full three-dimensional potential model
(\ref{eq:sumpot}).
The projection was computed by making an integration over the
$y$-coordinate, although in practice, given the symmetry of the crystal,
this was found to be equivalent to evaluating the full three-dimensional
potential at a given value of such coordinate.

\begin{table}
 \caption{\label{tab:pengvalues}
 Fitting parameters of the atomic scattering potentials obtained by
 Peng for Al \cite{peng1999electron}.}
 \centering
 \begin{tabular}{c|cc}
  \hline
  $k$& $a_k$ & $b_k$ \\
  \hline\hline
  1&0.3582&0.4529 \\
  2&0.9754&3.7745 \\
  3&2.6393&23.3862 \\
  4&1.9103&80.5019 \\
  \hline
 \end{tabular}
\end{table}

The initial wave function is a Gaussian wave packet, with
perpendicular incident  momentum $p_z = \sqrt{2m E_i}$, which reads as
\begin{eqnarray}
 \Psi(x,z,t=0) & = & \frac{1}{\sqrt{2\pi\sigma_x\sigma_z}}\ \!
   e^{-(x-x_0)^2/4\sigma_x^2+ip_xx/m}
 \nonumber \\
  & & \qquad \times e^{-(z-z_0)^2/4\sigma_z^2 +ip_z z/m} ,
\end{eqnarray}
centered at $(x_0,z_0)$, on top of the Al slab, and with $\sigma_x$ and
$\sigma_z$ accounting for the wave-packet width along parallel and
perpendicular directions.
When an initial tilting was considered, a rotation of the full wave
packet was performed instead of only adding the corresponding value of
the parallel component to the incident momentum.
This initial wave packet was mapped by a set of 90 Bohmian initial
conditions. For practical
purposes and to provide qualitative insight into the wave function
propagation, these initial conditions are distributed equidistantly
along three rows, each one spanning the interval $x_0\pm4\sigma_x$.
Equidistant initial positions have been chosen to show how each
particular region of the wave packet generates different dynamics in
spite of the fact that the initial phase, as specified by
$S(x,z,t=0)$, is the same for the whole of the wave function.
In other words, in this work each trajectory plays the role of an
individual hydrodynamical tracer \cite{sanz2012trajectory}.
In a Monte Carlo-like sampling, however, either the distribution of
initial conditions would follow the value of the corresponding initial
probability density or, in the case of an equidistant distribution, each
initial condition would be assigned a specific weight, according to the
value of the probability density in the close neighborhood of that
condition.

The effects on electron diffraction as a function of three
different incidence parameters have been investigated, namely the spot
size (beam width), accelerating voltage (incidence energy), and tilt
angle (incidence angle).
To this end, first, simulations were performed at normal incidence
with an accelerating voltage of 1~keV and two different spot
sizes, which are modeled in terms of two different wave-packet
widths along the $x$ (transversal) direction: $\sigma_x = a$
and $\sigma_x = 0.5a$. The spread in the $z$-direction was kept
constant at $\sigma_z=a$. Typically, TEM probe sizes can range from
0.1 to 1~nm, while probe sizes used in STEM are approximately 1~nm
\cite{williams2009transmission}. Given that the lattice parameter of
Al is near 0.4 nm, the probe sizes chosen in this work are in
accordance with those used in experiment
\cite{brandes_smithells_1998}. Subsequently, a comparison was
made between two tilt angles of the wave packet, at the negative
Bragg condition and at $-10^\circ$. These simulations were also
performed at 1~keV and the spot size was kept constant. Finally, two different
accelerating voltages, 0.1~keV and 6~keV, were investigated at
normal incidence and constant spot size.

%%%%%%%%%%%%%%%%%%%%%%%%%%%%%%%%%%%%%%%%%%%%%%%%%%%%%%%%%%%%%%%%%%%%%%%

\section{\label{sec3} Results and Discussion}

\begin{figure}[t]
 \centering
 \includegraphics[width=7.5cm]{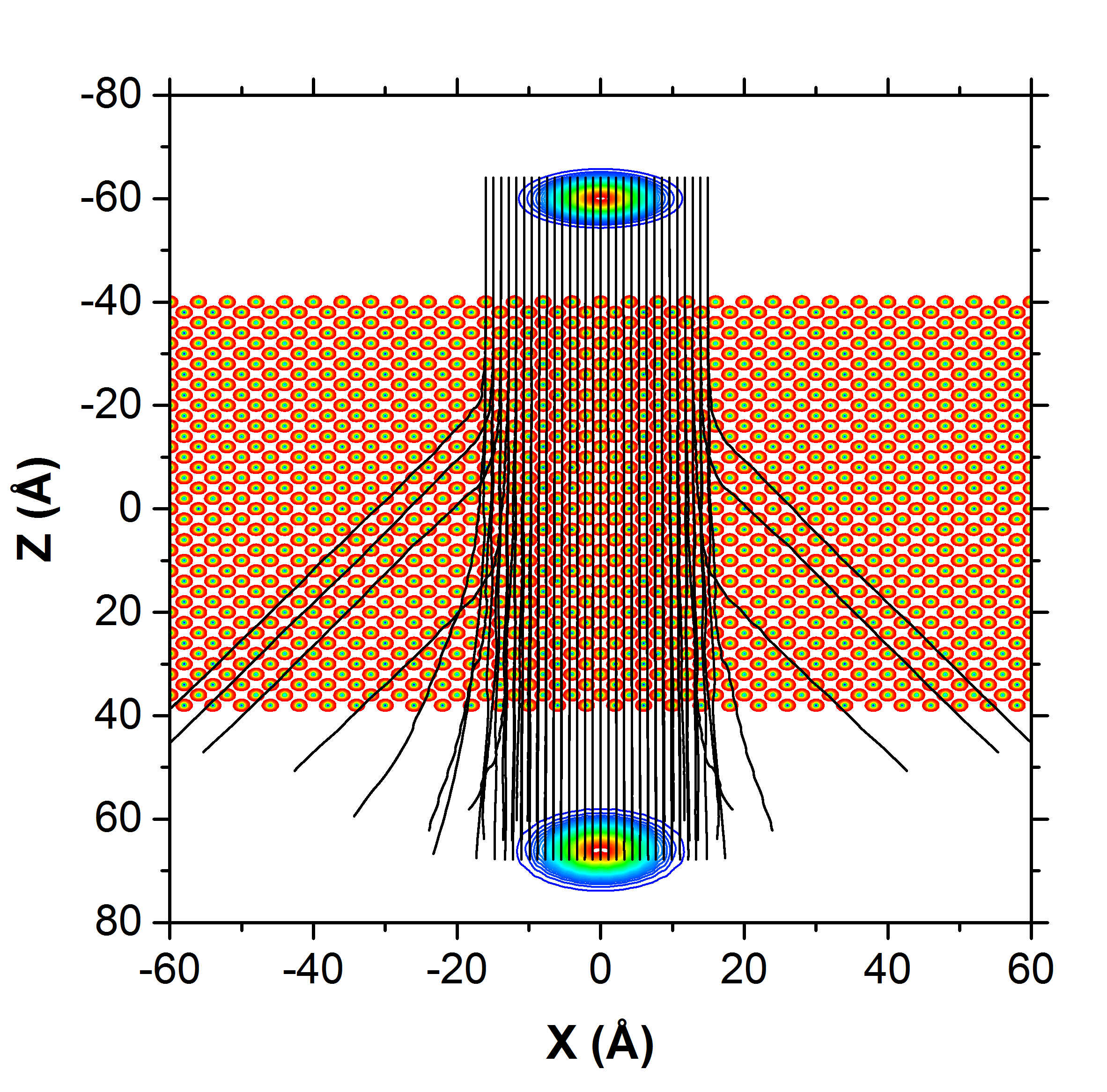}
 \caption{Swarm of Bohmian trajectories simulating the passage of
  an electron beam through an Al[100] crystal with an incidence energy
  $E=1$~keV and an angle equal to the Bragg angle
  $\theta_B = 2.773^\circ$.
  In this simulation, the electron-crystal interaction potential model
  consists of 40 atomic layers along the $z$ direction (an average over
  the $y$ direction, perpendicular to this page, has been considered
  in passing from the three to the two dimensional model).
  As a reference, the initial and final probability densities are also represented in terms of contour-plots.}
 \label{fig:Fig1}
\end{figure}

When the algorithm is initiated, the starting parameters are input
into the simulation. Next, the propagation of the wave function and,
subsequently, the trajectories are run until the wave function has
exited the Al thin film. Once the algorithm has terminated, the
configuration space outputs consist of the final probability density
of the wave function and coordinates of the trajectories at each
time step. The trajectories propagate from negative to
positive $z$ in accordance with EM conventions. A pictorial
representation of the simulation is depicted
in Fig.~\ref{fig:Fig1}.

\begin{figure}[t]
\centering
\includegraphics[width=7.5cm]{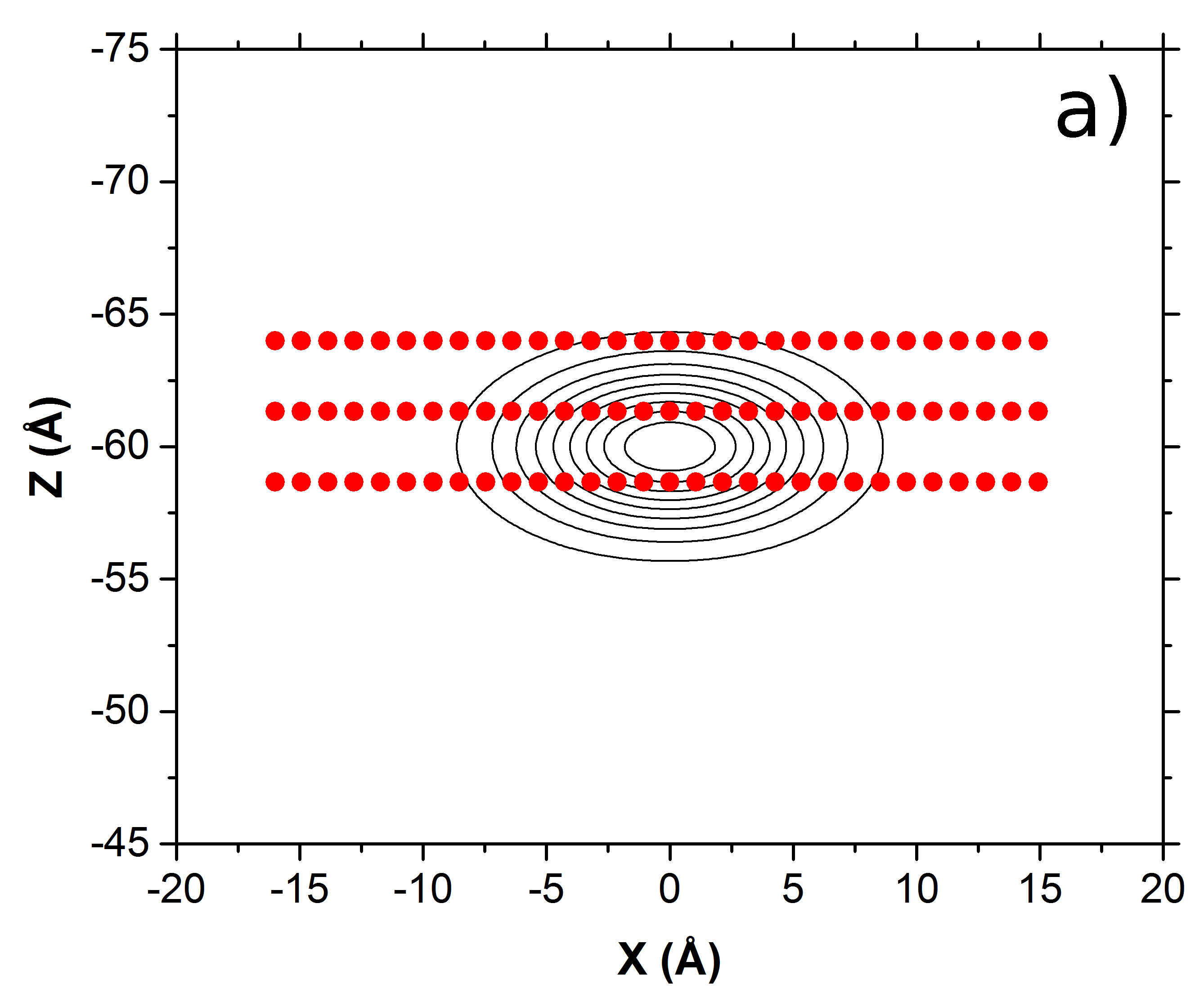}
\includegraphics[width=7.5cm]{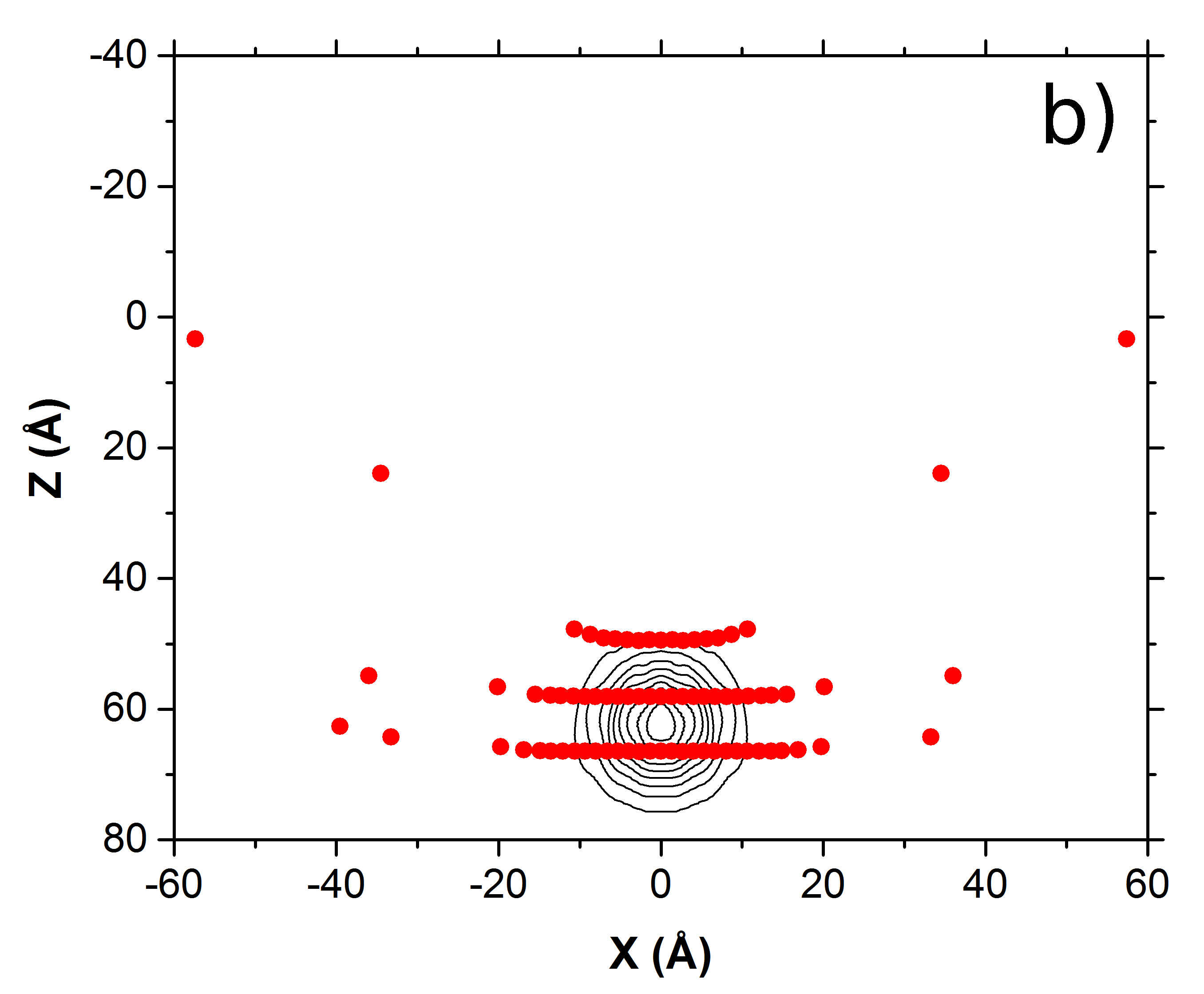}
 \caption{ (a) Initial positions for a swarm of Bohmian trajectories
 sampling the extension covered by the initial probability density
 (contour-plot).
 (b) Final positions reached by the trajectories started at the
 positions displayed in panel (a);
 the contour-plot of the final probability density is also displayed
 for the sake of comparison.}
\label{fig:Fig2}
\end{figure}

\begin{figure}[t]
 \centering
 \includegraphics[width=7.5cm]{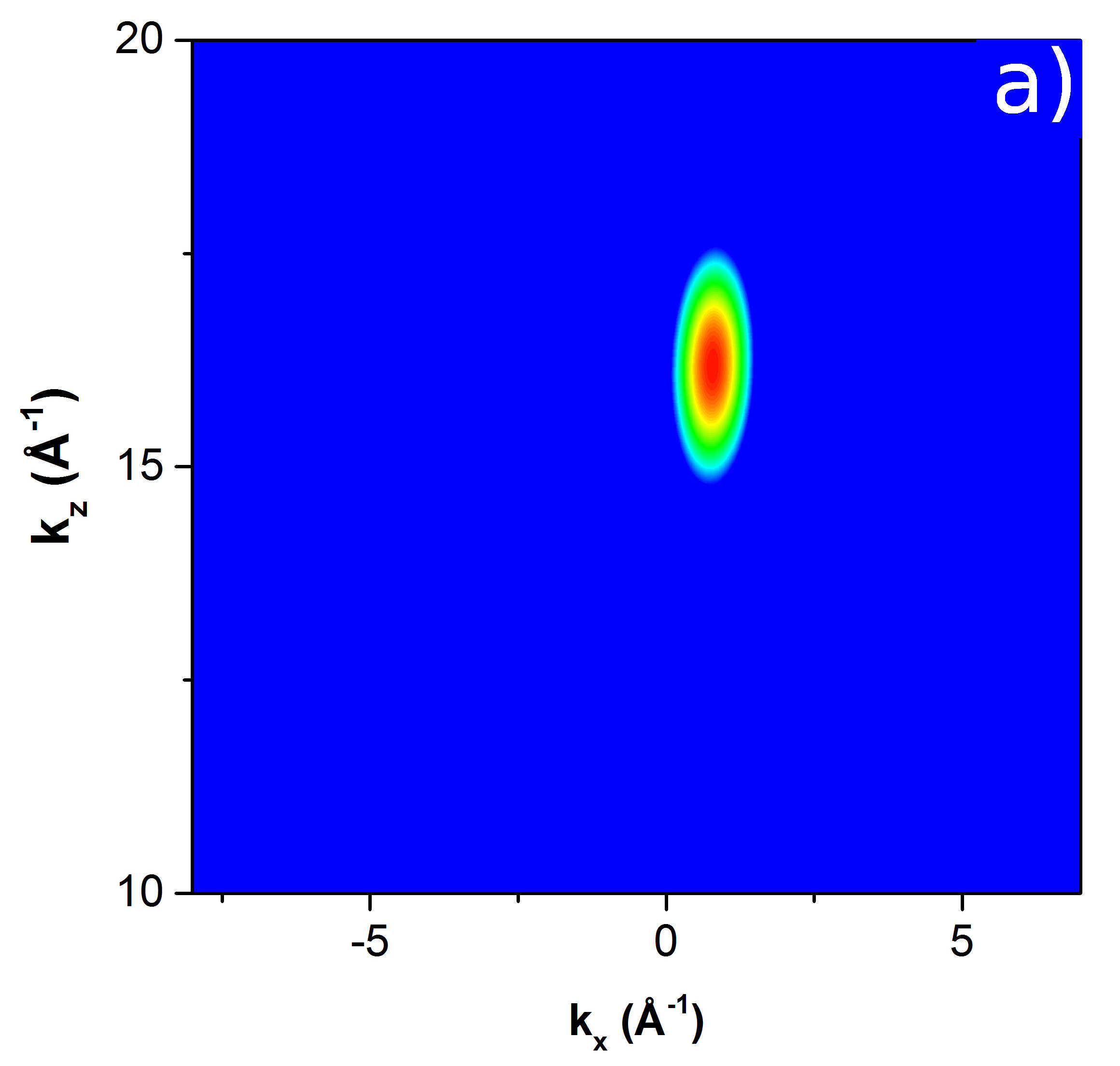}
 \includegraphics[width=7.5cm]{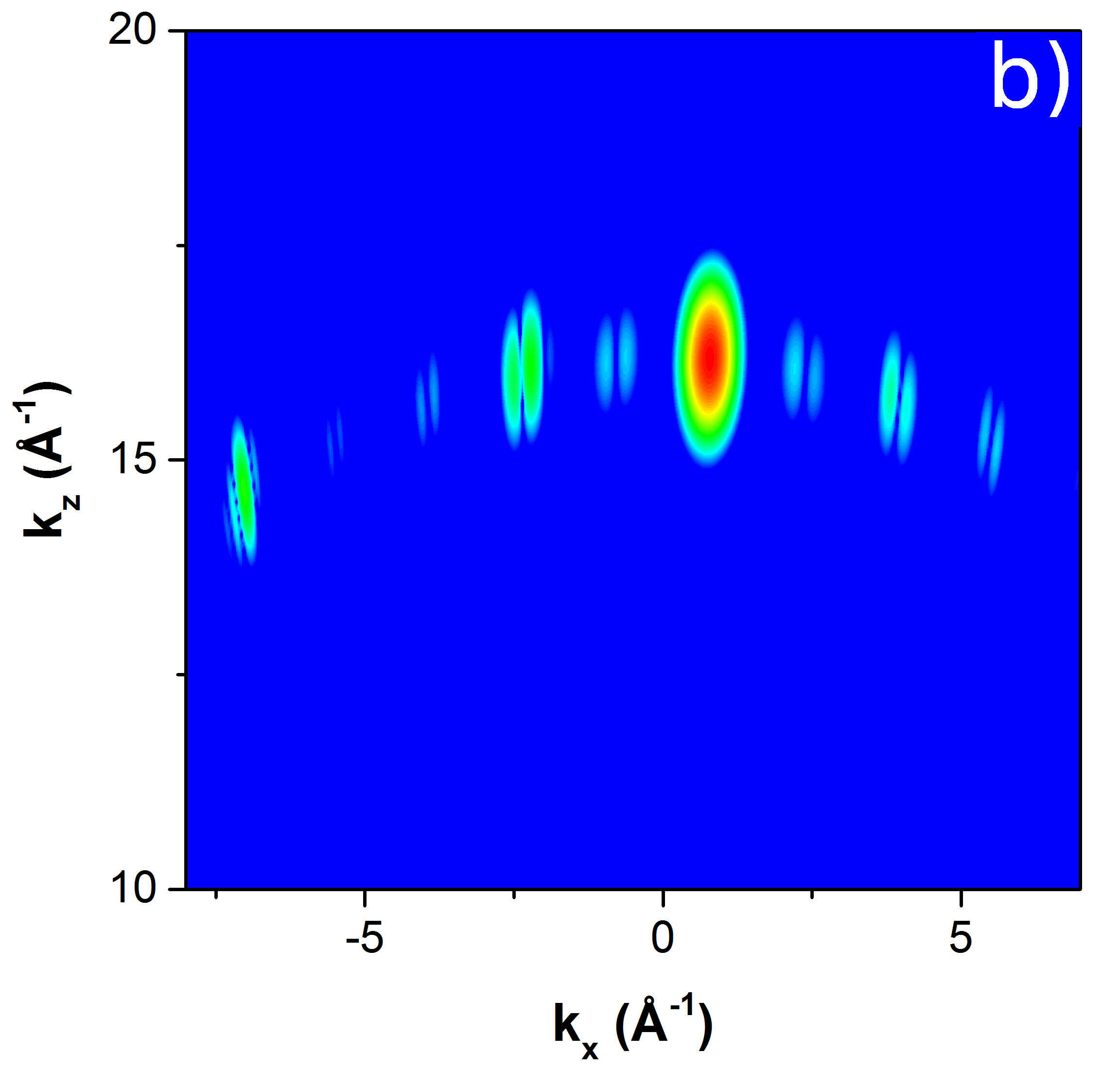}
 \caption{Initial (a) and final (b) probability densities in momentum
 space for an incidence at the Bragg angle $\theta_B \approx - 2.773^\circ$
 with an energy of 1~keV.}
 \label{fig:Fig3}
\end{figure}

The center of the initial wave function is at $x_0= 0$~\AA\ and
$z_0=-60$~\AA\ on the grid and it is propagated vertically with
positive momentum in $z$. After passing through the thin Al crystal,
a final probability density is retained and plotted. The
trajectories begin at the initial positions shown in
Fig.~\ref{fig:Fig2}(a) and their final positions are as in
Fig.~\ref{fig:Fig2}(b) once the simulation is terminated. Each
spot indicates the initial or final coordinates of a single
trajectory in configuration space.
Because of the laminarity displayed by the trajectories (the
so-called non-crossing property or rule of Bohmian mechanics
\cite{holland-bk,sanz2012trajectory}, although it is actually a
property of quantum mechanics itself \cite{sanz:JPA:2008}), there is
a direct correlation between the portion of the initial wave packet
covered by the Bohmian initial conditions and the final region reached
by the corresponding trajectories \cite{sanz:JPA:2008,sanz:JPA:2011},
even if we would know nothing about the actual paths.
This can be seen by inspecting the initial and final probability
densities, which have also been represented in Fig.~\ref{fig:Fig2} as
contour-plots, and provide a reference on the regions where
trajectories start and finalize.
Notice that the final positions of the trajectories contribute to the
different intensities that would be registered by a detector in a real
experiment, since the time-evolution of the swarm of trajectories
is in compliance with an exact quantum motion (this does not
necessarily mean that real electrons move along these paths, but only
their overall or averaged flow), unlike what happens with
classical-based methodologies.
These peaks represent diffraction spots, which are typically depicted
in reciprocal or momentum space. Figures~\ref{fig:Fig3}(a) and (b)
show the probability densities in momentum space at the start and
end of the simulation.

Since the initial wave function is a localized wave packet, there
is only a primary beam in momentum space. After propagation through
the Al single crystal, the beam is diffracted and spots can already
be seen in its momentum representation even though the wave function
is still in the Fresnel regime. The two adjacent peaks within a
single diffraction spot are near-field effects which show that
Fresnel diffraction is still apparent. In Fig.~\ref{fig:Fig3},
the wave packet was tilted to the Bragg condition in order to obtain
diffraction spots unique to the crystal. The difference in intensity
between certain diffraction spots is a consequence of the atomic
configuration of the projected FCC lattice. The alternating sequence
of atoms from one layer to another acts as two adjacent gratings
where the slits of the second grating are shifted by half a lattice
spacing compared to the first. The channels of the second layer
cause destructive interference of certain frequencies resulting in
less prominent diffraction intensities. However, the only
information that can be retained from this data is the intensity
distribution of the outgoing wave packet. The trajectories
themselves show which portions of the wave function are responsible
for these peaks and what occurs inside the material prior to
acquisition of this information. The following subsections evaluate
the evolution of quantum trajectories for each series of varying
initial conditions.

%%%%%%%%%%%%%%%%%%%%%%%%%%%%%%%%%%%%%%%%%%%%%%%%%%%%%%%%%%%%%%%%%%%%%%%

\subsection{\label{sec31} Probe size}

\begin{figure}[t]
 \centering
 \includegraphics[width=7.5cm]{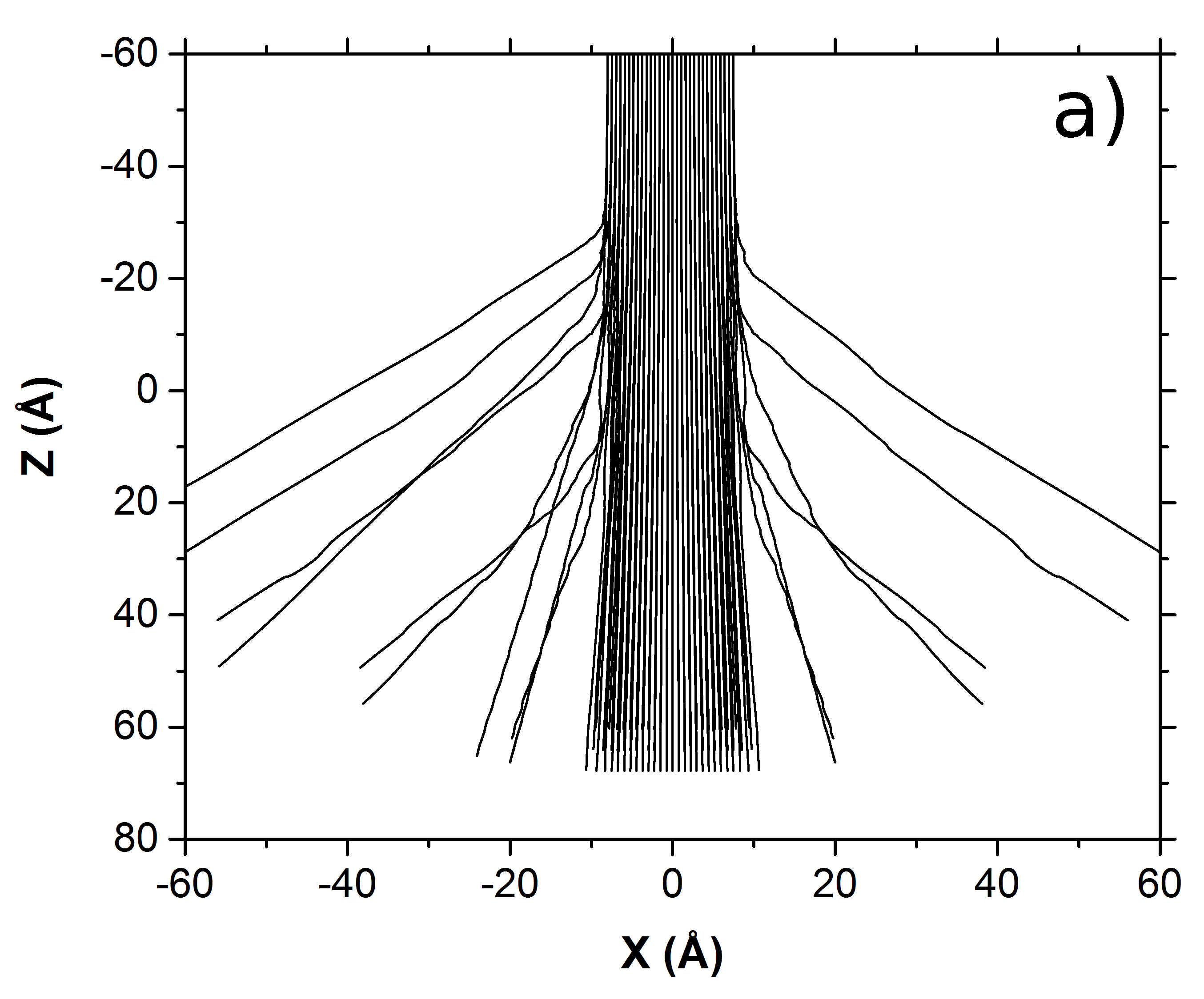}
 \includegraphics[width=7.5cm]{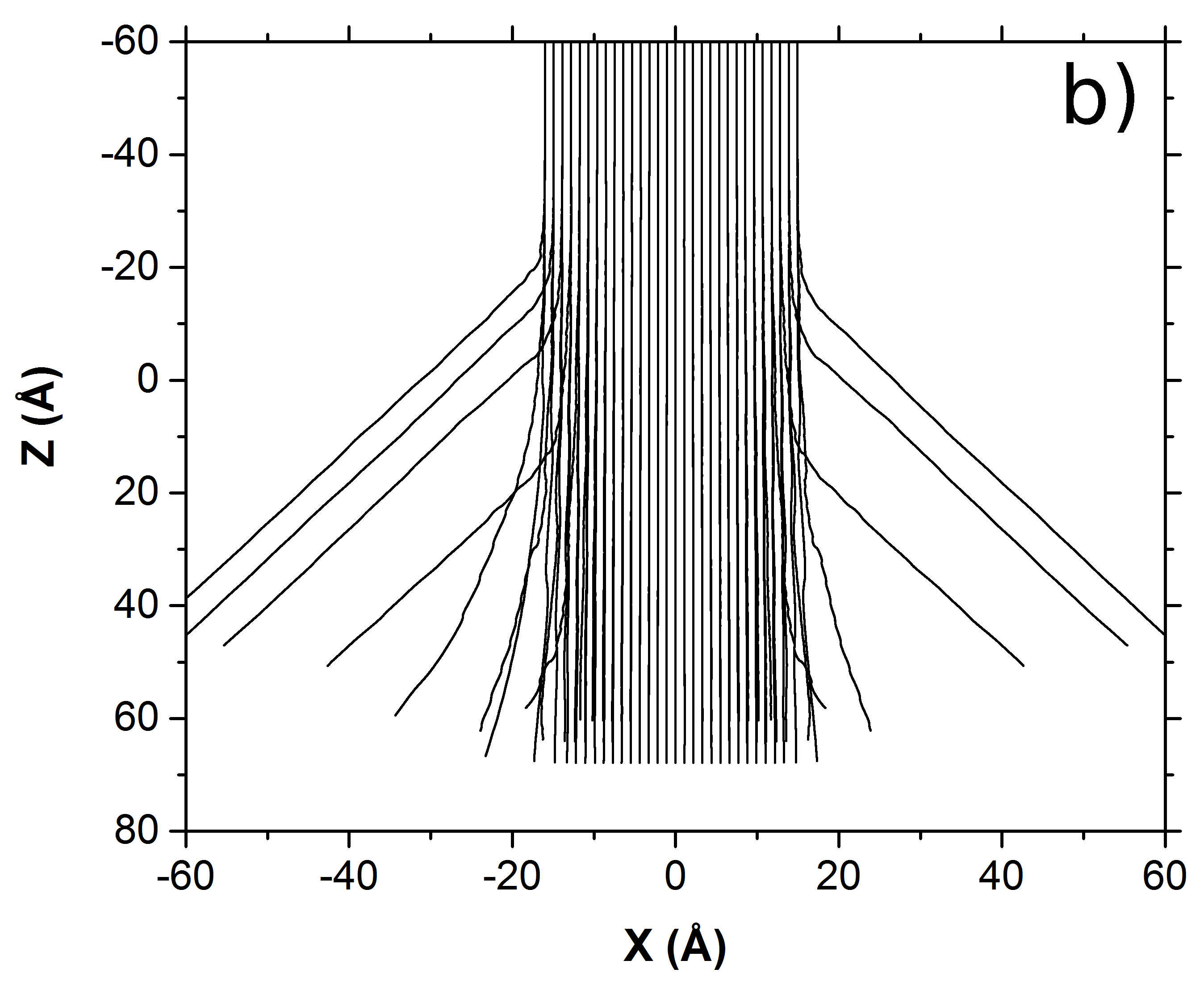}
 \caption{ Bohmian trajectories for perpendicular incidence
 ($\theta_i = 0^\circ$) with 1~keV and two different sizes of the
 probe: (a) $\sigma_x = 0.5a$ and (b) $\sigma_x = a$.}
 \label{fig:Fig4}
\end{figure}

Spot size is highly responsible for resolution in EM. A smaller spot
size increases resolution but, as a result, beam broadening can
become an issue \cite{Clifflorimer1981scattering}. The trajectories
at various spot sizes demonstrate what portion of the wave function
is responsible for such broadening and how this affects interactions
between the electrons and the material, again something arising
from the quantum non-crossing mentioned above.
Figure~\ref{fig:Fig4}
shows the propagation of quantum trajectories for two given probe
sizes, $\sigma_x=0.5a$ and $\sigma_x=a$.

Given the relatively thin material used with the given voltage of
1~keV, electrons mostly remain in the primary beam upon exit.
However, trajectories near $\pm3\sigma$ and $\pm4\sigma_x$  show the
portions of the wave function that are dispersed and contribute to
spreading. Comparing Figs.~\ref{fig:Fig4}(a) and (b) we find
that, as is well known from diffraction theory, the broader the
incident beam, the lesser the diffraction effects, and vice versa.
In terms of trajectories this effect manifests as a larger or
smaller amount of trajectories deviating from the incidence
direction during the travel through the material, which is followed by a large portion of the incoming swarm of trajectories.
Notice that, compared to the multislice technique, it is precisely
this deviation which cannot be described, since all trajectories are
forced to reach the same slab at the same time due to the
reparameterization of the longitudinal coordinate ($z$-direction) in
terms of the propagation time.
For any Gaussian wave packet, and specifically for an
electron beam, the final
spread of the wave packet is related to the initial spread by the
following,
\begin{equation}
 \label{eq:spread}
 \frac{\sigma}{\sigma_0}=\sqrt{1+\frac{\hbar^2t^2}{4m^2\sigma_0^4}} ,
\end{equation}
with $t$ equal to the time of propagation \cite{bohm1951quantum}. As
$\sigma_0$ decreases, the ratio of the initial and final spreads
increases exponentially until the Fraunhofer region is reached, at
which point the relationship becomes linear \cite{sanz:JPA:2008}.
This implies that beam
broadening of smaller probe sizes is increasingly greater than that
of larger probe sizes. The values of $\sigma/\sigma_0$ for
$\sigma_0=0.5a$ and $\sigma_0=a$ are 1.46 and 1.02 respectively.
Looking at the trajectories, taking the ratio of the average final
position in $x$ over the average initial position in $x$ for
trajectories that lie within 99\% of the Gaussian beam yields 1.42
for $\sigma_0=0.5a$ and 1.03 for $\sigma_0=a$. This demonstrates
that the numerical model is in agreement with analytical theories
and the drift of the trajectories is proportional to the wave packet
spreading. It is important to note that 99\% of the beam broadening
is approximately 3$\sigma$ from the mean, indicating that
trajectories outside of this range are unlikely to contribute to the
broadening. The minimal broadening calculated for the simulation
where $\sigma_0=a$ can be visualized in Fig.~\ref{fig:Fig4}(b)
where the outlying trajectories do not contribute to the spreading
and the trajectories within 99\% of beam broadening remain within
the same range upon exit. On the other hand, with $\sigma_0=0.5a$,
the beam spreads by nearly half of the initial probe size regardless
of the trajectories outside 99\%. In practice, this may cause issues
in resolution and data analysis. In all simulations, the initial
positions of the trajectories were chosen deterministically whereas
the initial wave function has the form of a Gaussian distribution.
Therefore, to quantify the average drift of electrons in the beam, a
weighted average of the distances between the initial and final
positions in $x$ must be used, where the weights are the integral of
the initial Gaussian about each starting point. This ensures that
the appropriate contributions of each trajectory representative of
the wave function are considered. With this, the average drift of
trajectories in Fig.~\ref{fig:Fig4}(b) is 0.0043~\AA, while that
from Fig.~\ref{fig:Fig4}(a) is 0.080~\AA. A greater drift, and
as a result high beam broadening, is still seen at a smaller probe
size.

%%%%%%%%%%%%%%%%%%%%%%%%%%%%%%%%%%%%%%%%%%%%%%%%%%%%%%%%%%%%%%%%%%%%%%%

\subsection{\label{sec33} Tilt angle}

\begin{figure}[t]
 \includegraphics[width=7.5cm]{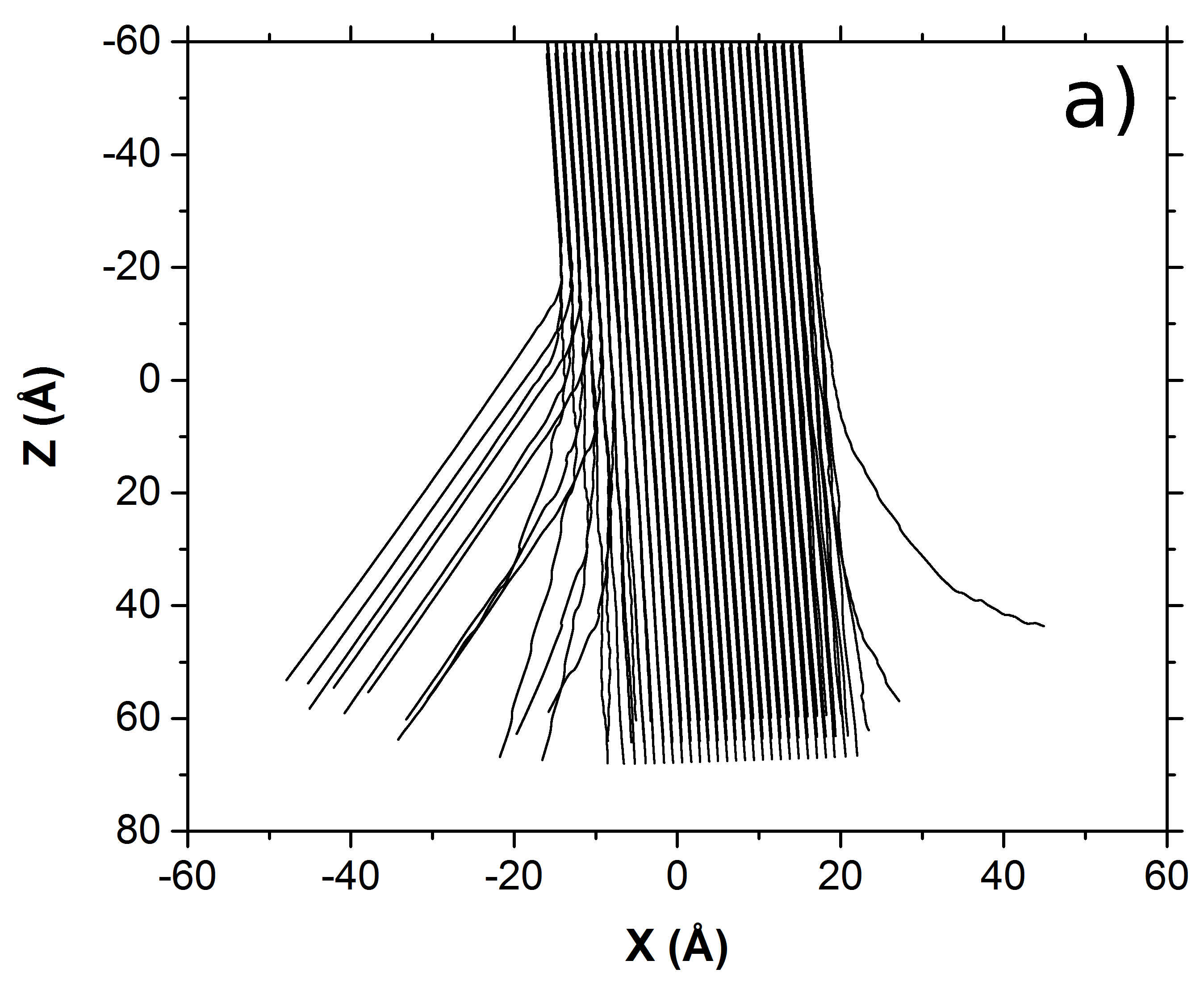}
 \includegraphics[width=7.5cm]{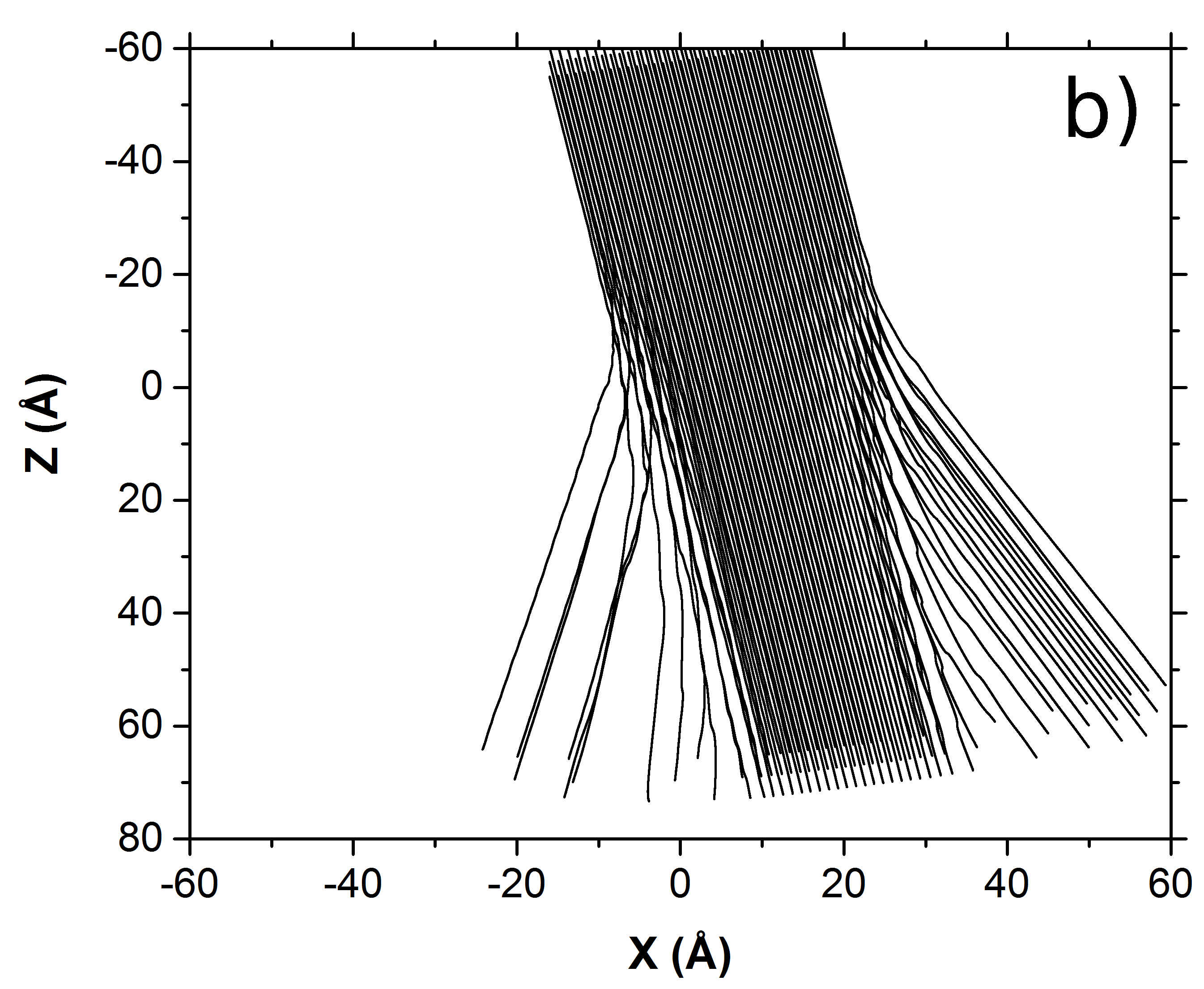}
 \caption{Bohmian trajectories associated with a probe size
 $\sigma_x = a$ for an accelerating voltage of 1~keV and two different
 incidence angles: (a) a Bragg angle $\theta_i = \theta_B \approx - 2.773^\circ$
 and (b) $\theta_i = -10^\circ$.}
 \label{fig:Fig5}
\end{figure}

\begin{figure}[t]
 \centering
 \includegraphics[width=7.5cm]{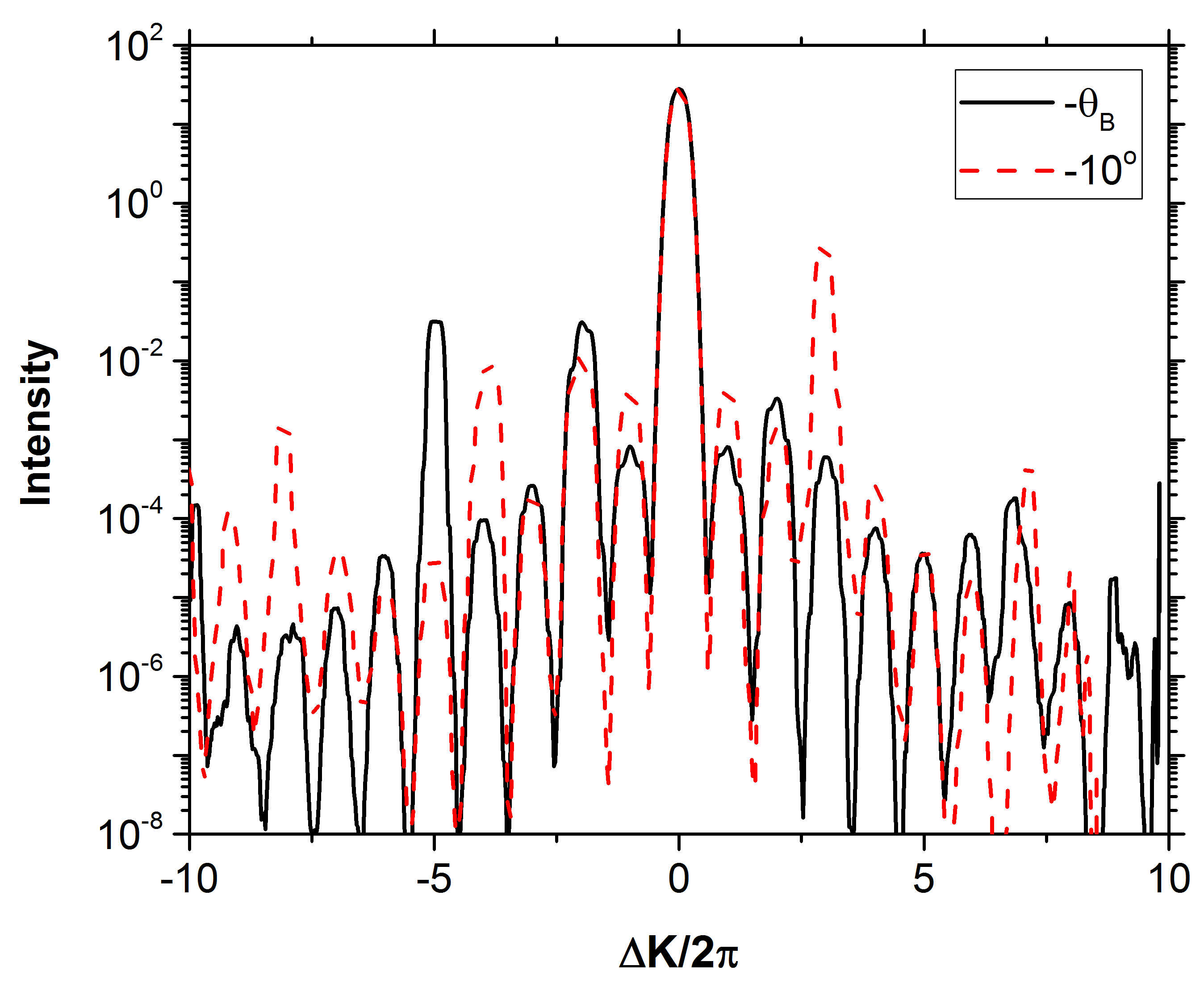}
 \caption{Diffraction intensity distribution as a function of the
 parallel momentum transferred for two different incidence angles: a Bragg angle $\theta_i = \theta_B \approx - 2.773^\circ$ and $\theta_i = -10^\circ$.}
 \label{fig:Fig6}
\end{figure}

An important advantage of using this method for simulating
an electron beam through a thin film is the possibility of
reproducing various tilt angles, as shown in
Figs.~\ref{fig:Fig5}(a) and (b).
At 1~keV, the Bragg angle is $\theta_B \approx -2.773^\circ$. This
is the condition at which diffraction spots unique to the crystal
orientation will appear in the Fraunhofer regime. However, they are
still present in Fresnel diffraction and can provide information
about the material. Higher tilt angles may be used in experiments
such as transmission electron forward scattering diffraction
(t-EFSD) where Kikuchi patterns can be imaged \cite{JMI:JMI12007}.
Here, any tilt angle can be simulated
by rotating the initial wave packet and attributing the appropriate
momenta to both spatial components. Then, the portions of the wave
function that are either deviated, diffracted or even backscattered
by the Al crystal can be distinguished through the trajectories. At
a $-10^\circ$ tilt, although the Bragg condition is not satisfied,
the tail ends of the wave function are still highly deviated. This
could be due to dispersion of the wave function as a result of the
high incidence angle. As mentioned previously, the thin specimen
acts as a grating. The dispersion, $D$, of a grating is calculated
by the following.
\begin{equation}
 D = \frac{|m|}{b\cos\theta_0} ,
 \label{eq:dispersion}
\end{equation}
where $b$ is the spacing of the grating, $m$ is the spectral order,
and $\theta_0$ is the incidence angle \cite{elmore1969physics}. As
$\theta_0$ increases, approaching $\pi/2$, the denominator of
Eq.~\ref{eq:dispersion} tends to zero and the dispersion increases.
This can be observed in Fig.~\ref{fig:Fig5}(b) by the low
densities of coupled trajectories. At the Bragg condition, the
portions of the wave function which contribute to the first and
second diffraction beams can be discerned by high densities of the
coupled portion of trajectories diffracted away from the primary
beam. Given the crystal structure, contributions to the second
diffraction peak away from the primary beam are greater than those
to the first. This is evident by the momentum representation of the
wave function in Fig.~\ref{fig:Fig3}(b) and is further verified
by calculation of the diffraction intensity distribution of the
final wave function computed through the S-matrix,
Fig.~\ref{fig:Fig6}.

While the primary beam has the highest intensity, the intensity of
the second peak is also considerably high. Again, the forbidden
reflection of the first diffraction spot is due to the alternating
sequence of atomic rows in the crystal. The peak at $\Delta
K/2\pi=-2$ is more intense then that at $\Delta K/2\pi=+2$ because
of the negative tilt to the Bragg angle. Compared to the intensity
distribution of the tilt to $-10^\circ$, whose peaks are convoluted
indicating incoherent scattering and high dispersion of the beam.
The trajectories themselves demonstrate precisely which portion of
the wave function contributes to each intensity.
%%%%%%%%%%%%%%%%%%%%%%%%%%%%%%%%%%%%%%%%%%%%%%%%%%%%%%%%%%%%%%%%%%%%%%%

\subsection{\label{sec32} Accelerating voltage}

A beam of electrons interacts very differently with a material
depending on the accelerating voltage. While little to no
interactions may be seen through thin films at high energies,
effects such as backscattering and high angle collisions may be
observed at low energies. In EM, the choice of accelerating voltage
is highly dependent on the phenomenon being characterized. To reach certain
ionization energy levels, accelerating voltages near 20-30~keV are
often used for SEM of bulk specimens. On
the other hand, these energies are still considerably lower than
those used in a TEM. Lower energies may be used for high resolution
imaging of small structures such as grains within Al-Li alloys
\cite{MAM:8788426}. Although these studies are done on bulk
materials, it is interesting to investigate the effects of low
energies on thin films to make parallels between the two sample
types. Therefore, trajectories computed for different accelerating
voltages can show at which point the time-dependence of the wave
function may or may not be a factor.

\begin{figure}[t]
 \centering
 \includegraphics[width=7.5cm]{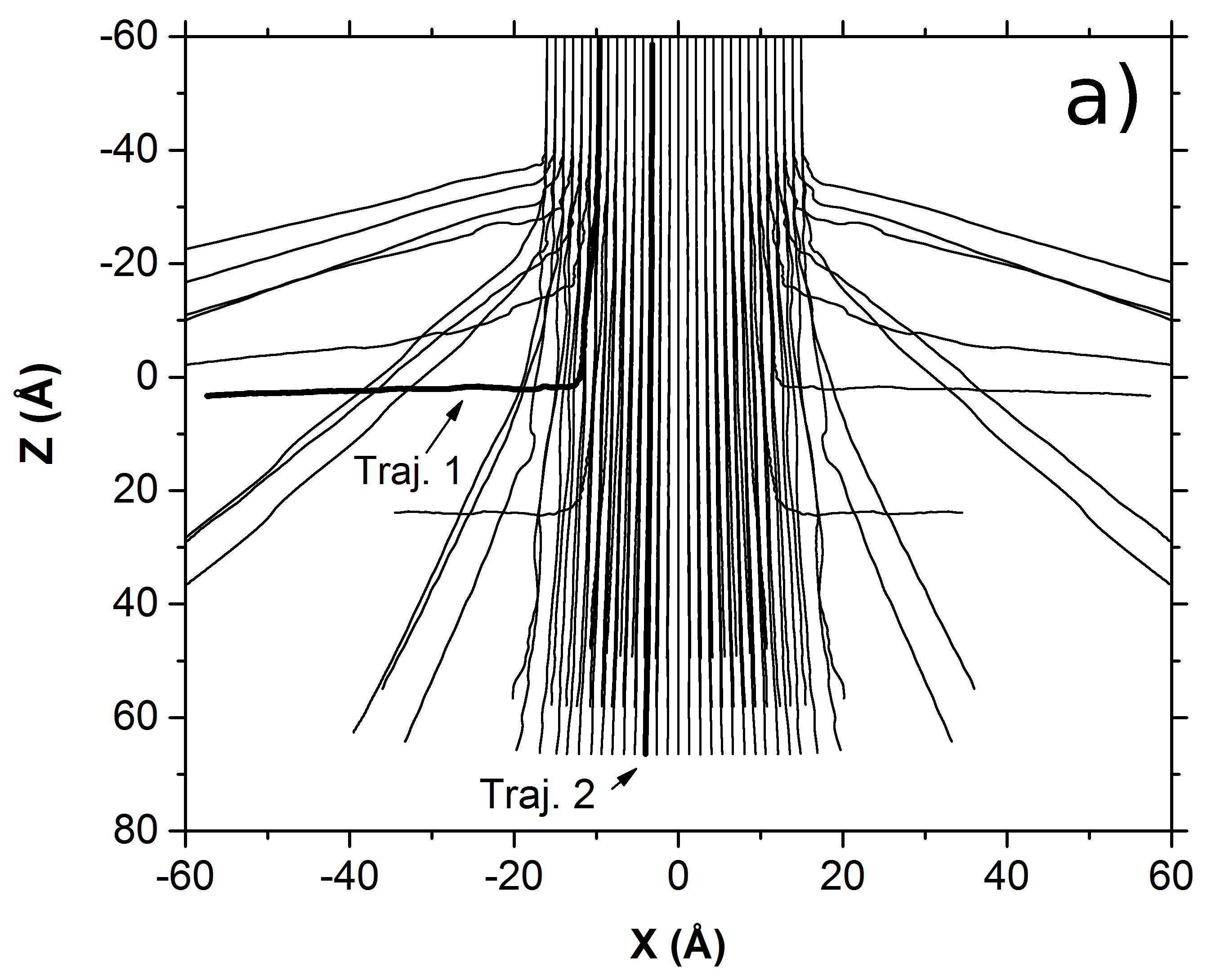}
 \includegraphics[width=7.5cm]{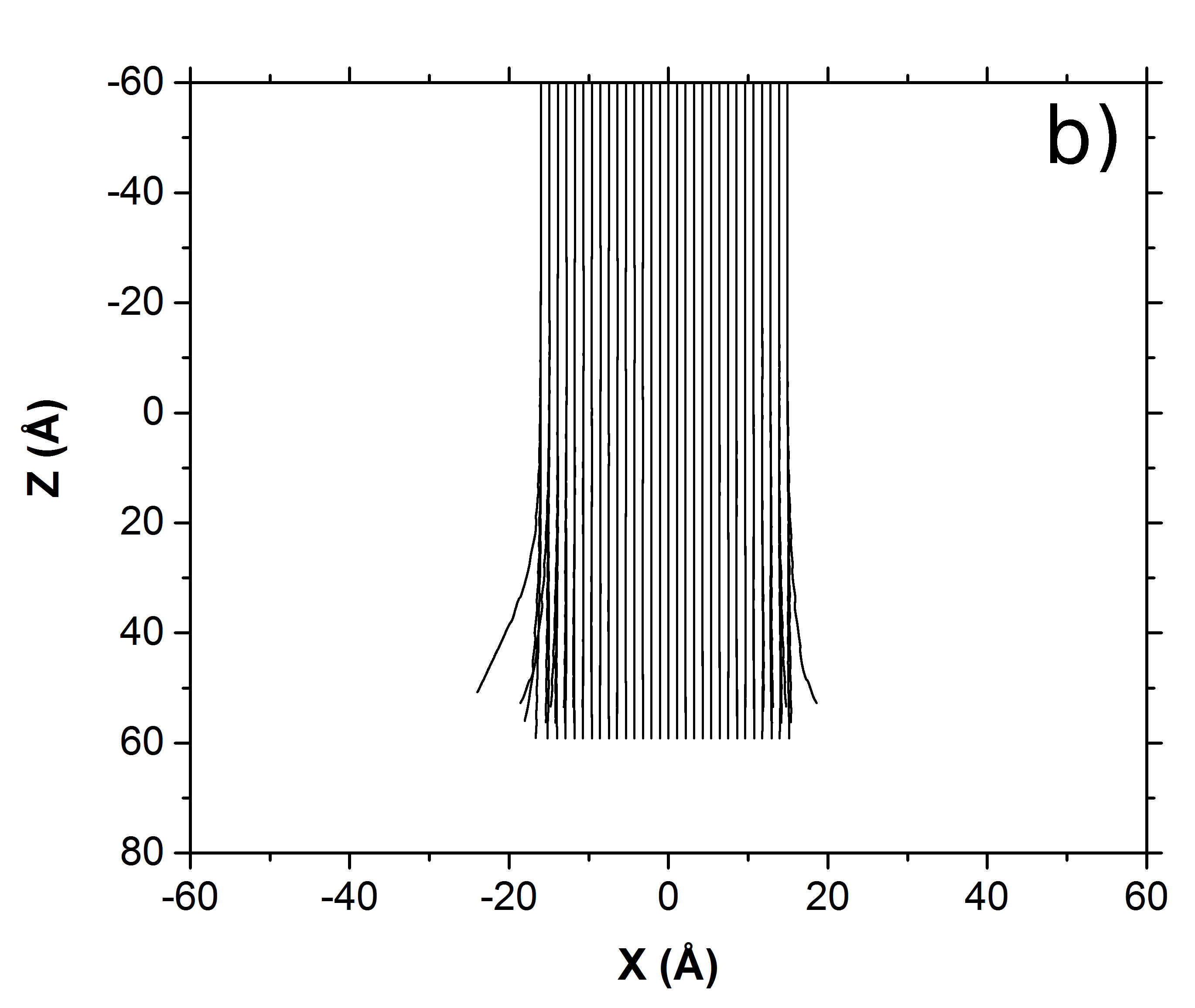}
 \caption{ Bohmian trajectories associated with a probe size
 $\sigma_x = a$ for perpendicular incidence ($\theta_i = 0^\circ$) and
 two different values of the accelerating voltage: (a) 0.1~keV and (b)
 6~keV.}
 \label{fig:Fig7}
\end{figure}

Figure~\ref{fig:Fig7}(b) shows trajectories computed for a wave
packet at an energy of 6~keV. It can be seen that there is very
little change in the trajectory paths with a few single paths that
could be deviated in the far-field. In such a case, the result
resembles the trajectories obtained by Zhang {\it et al.}, where it
is assumed that the guiding equations are independent of time and
only a constant flow is calculated \cite{Zhang2015JMicro}. An
accelerating voltage of 6~keV is sufficient to show complete
transmission of the beam because of the small thickness of the
material simulated. There could in fact be scattering at this energy
with thicker materials and scattering will always be present in bulk
materials. Therefore, a time-independent approach cannot be used if
the desire is to investigate various energies or sample thicknesses.
In a 20 atom layer sample, 0.1~keV is sufficient to show the presence
of backscattering, as in Fig.~\ref{fig:Fig7}(a), where collisions
can cause particles to remain inside the material or even exit the
material at the surface of incidence. This is not normally seen in
thin materials, however the trajectories at the tail end of the wave
function in Fig.~\ref{fig:Fig7}(a) show a complete transfer of
momentum from the $z$-direction to the $x$-direction indicating that
in a very small percentage of the incident electrons, some high
angle scattering may occur. In this case, the kinetic energy of the
trajectories varies considerably compared to trajectories which do
not suffer such scattering.

\begin{figure}[t]
 \centering
 \includegraphics[width=7.5cm]{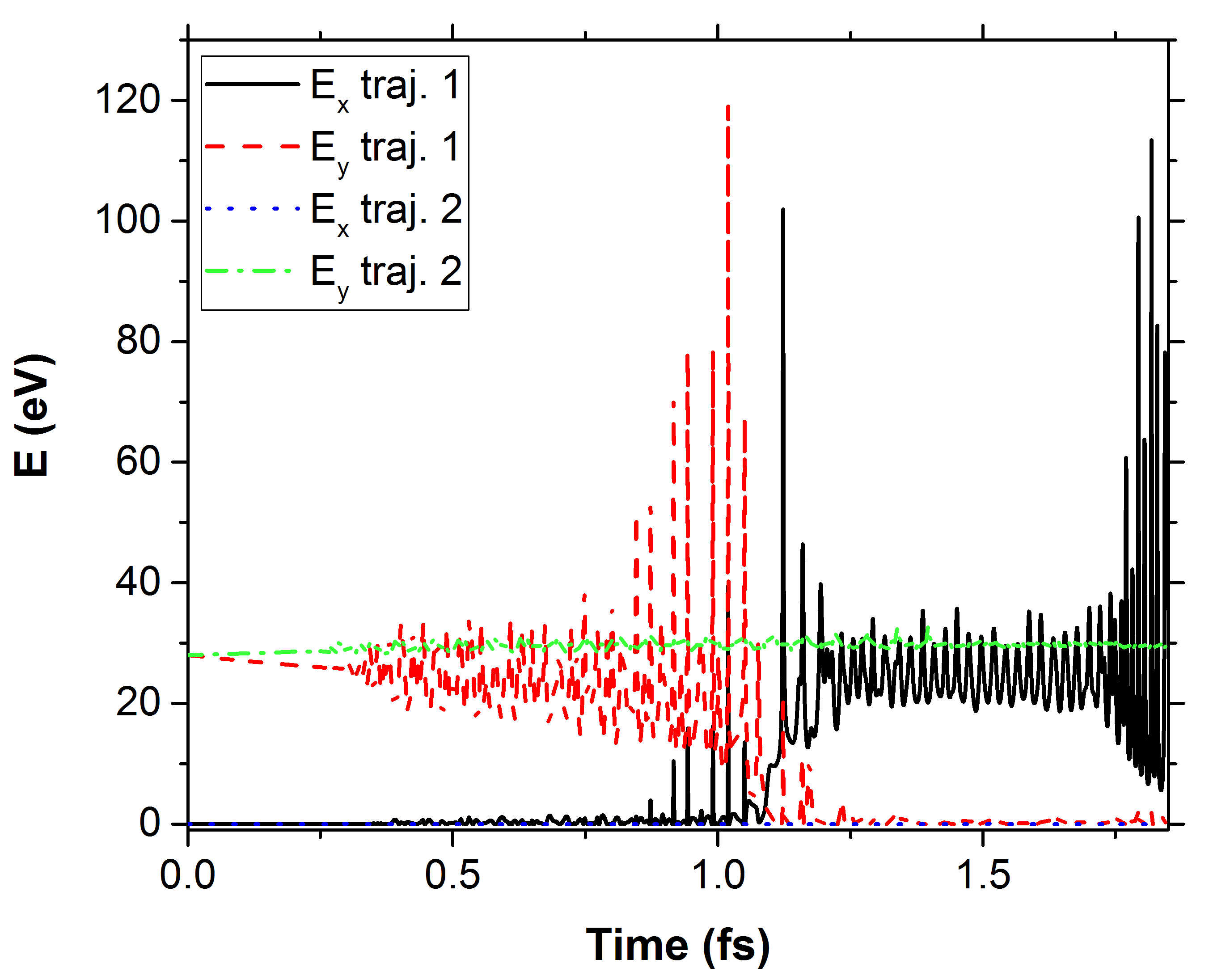}
 \caption{Perpendicular ($z$) and parallel ($x$) components of the
  kinetic energy for a Bohmian trajectory undergoing a high angle
  deviation (traj.~1) and another remaining in the primary beam
  (traj.~2), accelerated by a voltage of 0.1~keV.}
 \label{fig:Fig8}
\end{figure}

\begin{figure}[t]
 \centering
 \includegraphics[width=7.5cm]{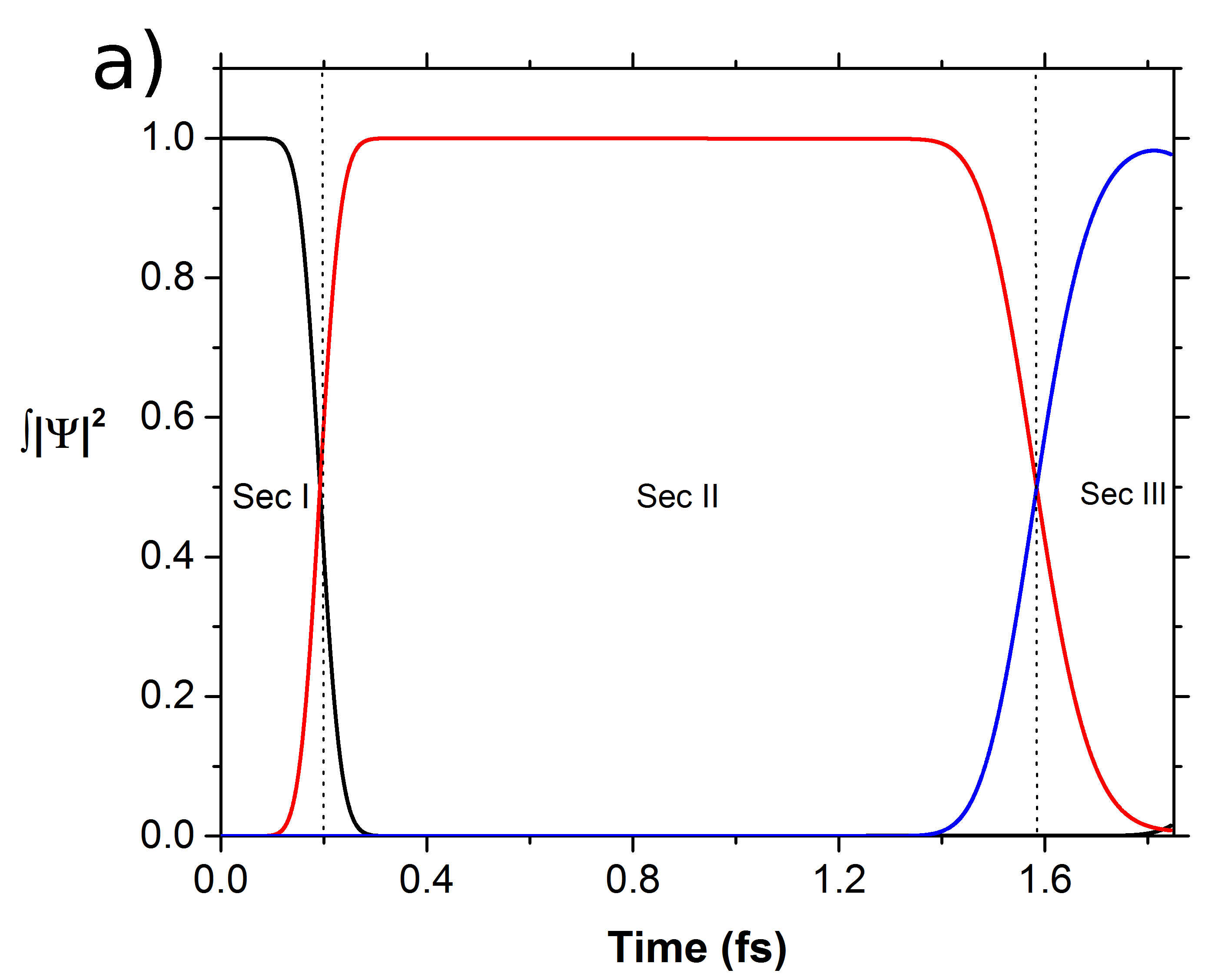}
 \includegraphics[width=7.5cm]{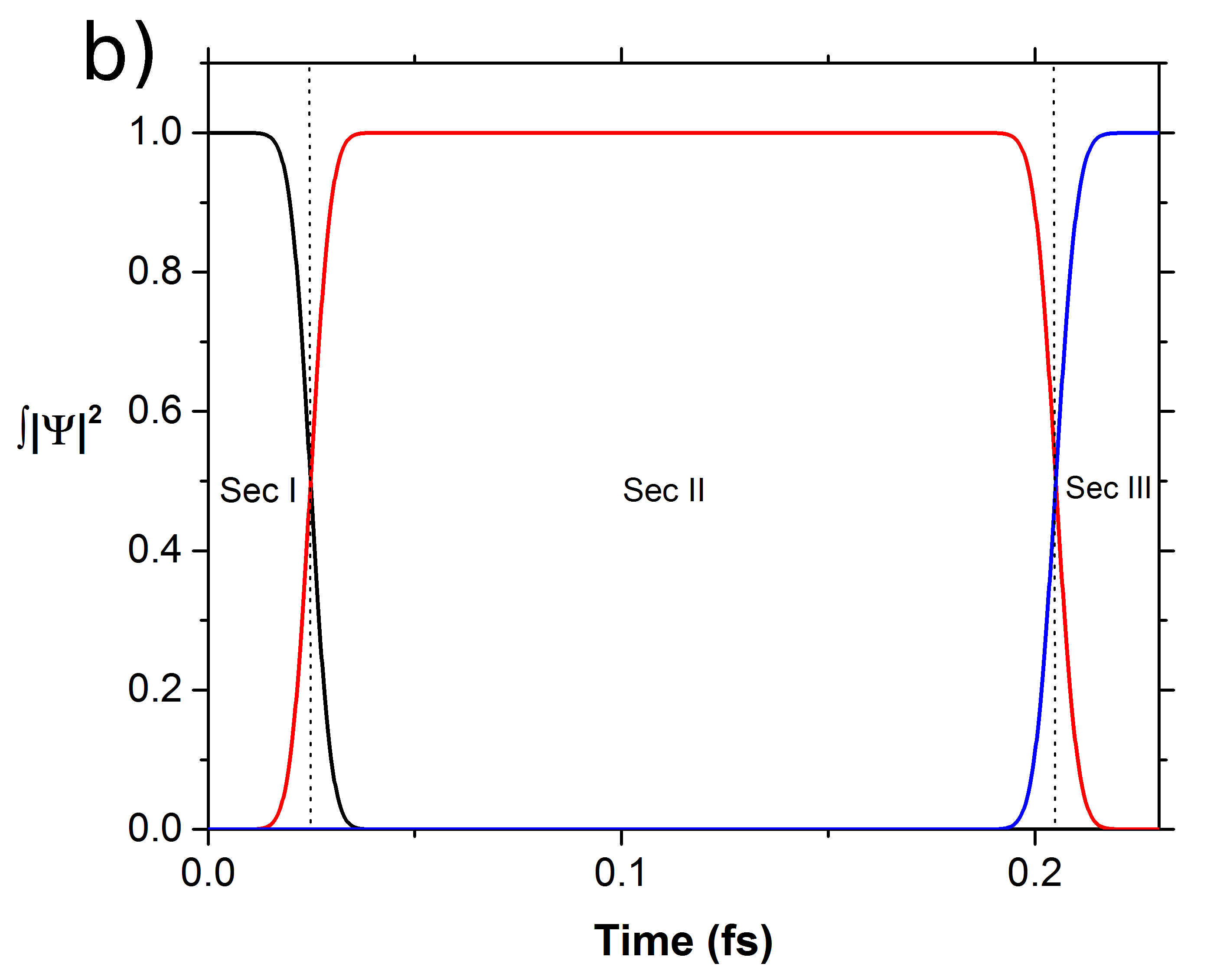}
 \caption{Evolution of the restricted probabilities, as given by
  Eq.~(\ref{rest}), before (I), inside
 (II) and after (III) the material for two different accelerating
 voltages: (a) 0.100~keV and (b) 6~keV.}
 \label{fig:Fig9}
\end{figure}

This is shown in Fig.~\ref{fig:Fig8}, where the $x$ and $z$
components of the kinetic energy of a trajectory that is highly
deviated are compared to those of one that remains in the primary
beam. Trajectories 1 and 2 are indicated in
Fig.~\ref{fig:Fig7}(a). The values are calculated using the
momentum, $p$, of the trajectories as $E_K=p^2/2m$.
While both components of trajectory 2 remain relatively constant,
with negligible kinetic energy in the $x$-direction, there are high
fluctuations in the energy components of trajectory 1. Near 1 fs,
there is a transfer of energy between the $z$ and $x$ components.
This would indicate the point at which the trajectory's path
deviates towards the $x$-direction. The high fluctuations throughout
time in the kinetic energy of trajectory 1 can be attributed to the
effects of the crystal potential on this part of the wave function.
These backscattering effects could not be reproduced with reduced
trajectories which use the multislice algorithm to obtain the
corresponding wave function. Conversely, a full physical picture of
the electron beam interaction with the material is easily observed
through this quantum trajectory method. The difference in
propagation between the two accelerating voltages can also be shown
by comparing the restricted probabilities \cite{sanz:JPA:2011},
i.e., the space integral of the probability density within a
given space region, as given by Eq.~(\ref{rest}).
Thus, in our case, as mentioned in Sec.~\ref{sec2}, the regions of interest are delimited by the material
slab: before the material (I), inside the material (II), and after
the material (III), for both energies as shown in
Fig.~\ref{fig:Fig9}.

At 6~keV, the probability density is completely transferred
to region II after a certain period of time and subsequently transferred to region III at a later time, where these transitions
occur very rapidly. However, at 0.1~keV, even though most of the
probability density is transferred from region II to region
III, this transition is much slower indicating a longer residence
time inside the material. With a thicker sample material, this
transition may become even slower and the curve associated with
region II may not ever reach zero.

%%%%%%%%%%%%%%%%%%%%%%%%%%%%%%%%%%%%%%%%%%%%%%%%%%%%%%%%%%%%%%%%%%%%%%%

\section{\label{sec4} Conclusions}

Simulations of an electron beam through a thin Al single crystal
were performed using the theory of Bohmian mechanics. The wave
function of a 2D electron beam passing through the material was
simulated and quantum trajectories were computed to quantitatively
describe the propagation of the wave function.
Specifically, this work performs 2D simulations in order to
provide preliminary demonstrations of the advantages of this
time-dependent method of computing quantum trajectories while retaining
some computational efficiency. Extensions into a 3D time-dependent
situation or slicing approach are also possible and are investigated in
work currently under development. Furthermore, if propagation is continued into the far-field, Fraunhofer diffraction is distinguishable in configuration space when computed by a free particle time evolution of the wave function following its transmission through the material.

Thus, it has been shown that phenomena that are not distinguishable or unambiguous by current diffraction simulation techniques, such as backscattering and
the possibility of high angle collisions, can easily be observed
by the methodology here employed.
It has also been shown that initial
conditions, such as low accelerating voltages and high tilt angles by
rotation of the initial wave function, can be introduced in the
numerical simulation of TEM data analysis, while they are out of scope
in other currently existing and widely used methods in the literature.

%%%%%%%%%%%%%%%%%%%%%%%%%%%%%%%%%%%%%%%%%%%%%%%%%%%%%%%%%%%%%%%%%%%%%%%

\acknowledgments

R.G. and S.R. would like to acknowledge the Aluminum Research
Group (REGAL) for their financial support.

%%%%%%%%%%%%%%%%%%%%%%%%%%%%%%%%%%%%%%%%%%%%%%%%%%%%%%%%%%%%%%%%%%%%%%%

%\bibliographystyle{unsrt}
%\bibliography{bibliography}{}
%
\end{document}